\documentclass[conference]{IEEEtran}

\ifCLASSINFOpdf
   \usepackage[pdftex]{graphicx}
   \graphicspath{{./}}
   \DeclareGraphicsExtensions{.pdf,.jpeg,.png}
\else
\fi
\usepackage[cmex10]{amsmath}
\begin{document}
%
\title{Influence of sample geometry on inductive damping measurement methods}



%
\author{\IEEEauthorblockN{
N. Liebing\IEEEauthorrefmark{1},
S. Serrano-Guisan\IEEEauthorrefmark{1},
A. Caprile\IEEEauthorrefmark{2}\IEEEauthorrefmark{3}, 
E. S. Olivetti\IEEEauthorrefmark{2},
F. Celegato\IEEEauthorrefmark{2},
M. Pasquale\IEEEauthorrefmark{2},
A. M\"uller\IEEEauthorrefmark{1}\\ and
H.W. Schumacher\IEEEauthorrefmark{1}
}
\IEEEauthorblockA{\IEEEauthorrefmark{1}Physikalisch-Technische Bundesanstalt, Bundesallee 100, D-38159 Braunschweig, Germany}
\IEEEauthorblockA{\IEEEauthorrefmark{2}Instituto Nazionale di Ricerca Metrologica, Strada della Cacce 91, 10135 Torino, Italy}
\IEEEauthorblockA{\IEEEauthorrefmark{3}Politecnico di Torino, Dipartimento di Fisica, Corso Duca degli Abruzzi 24, 10129 Torino, Italy}
}

\IEEEspecialpapernotice{This document is the manuscript of a work that appeared in final form in IEEE Transactions on Magnetics, vol. 47(10),2502-2504, 2011, DOI: 10.1109/TMAG.2011.2155637, © IEEE after technical editing by the publisher. To access the final edited and published work see http://ieeexplore.ieee.org }

\maketitle

\begin{abstract}
We study the precession frequency and effective damping of patterned permalloy thin films of different geometry using integrated inductive test structures.  The test structures consist of coplanar wave guides fabricated onto patterned permalloy stripes of different geometry. The width, length and position of the permalloy stripe with respect to the center conductor of the wave guide are varied. The precession frequency and effective  damping of the different devices is derived by inductive measurements in time and frequency domain in in-plane magnetic fields. While the precession frequencies do not reveal a significant dependence on the sample geometry we find a decrease of the measured damping with increasing width of the permalloy centered underneath the center conductor of the coplanar wave guide. We attribute this effect to an additional damping contribution due to inhomogeneous line broadening at the edges of the permalloy stripes which does not contribute to the inductive signal provided the permalloy stripe is wider than the center conductor. Consequences for inductive determination of the effective damping using such integrated reference samples are discussed.
\end{abstract}


%
\IEEEpeerreviewmaketitle

\section{Introduction}
Inductive measurement methods, like pulsed inductive microwave magnetometry (PIMM) \cite{Silva1999} and two-port network analyzer ferromagnetic resonance measurements (VNA-FMR) \cite{Bilzer2007}, using a coplanar wave-guide (CPW) are fast and reliable techniques to measure dynamical properties of magnetic thin films and nanostructures. To establish these technique as a standard to perform traceable measurements it is necessary to understand the influence of different parameters on the measurement results. In most measurements an unpatterned magnetic thin film is placed on top of a CPW to measure its dynamic properties. However in this case the exact geometry of film and CPW could change e.g. due to absoprtion of a thin water film leading to an extrinsic increase of the measured damping \cite{Pasquale2010}.

Here, we use a broadband microwave setup to study the magnetization dynamics in the frequency and time domain of thin permalloy rectangular thin film samples as a function of the film geometry. To obtain a well defined experimental geometry integrated samples are used where CPW is directly patterned onto the patterned magnetic the film. The influence of the sample geometry on the measured dynamic properties is discussed.

\section{Samples and experimental setup}
\begin{figure}[!t]
\centering
\includegraphics[width=1.0\linewidth]{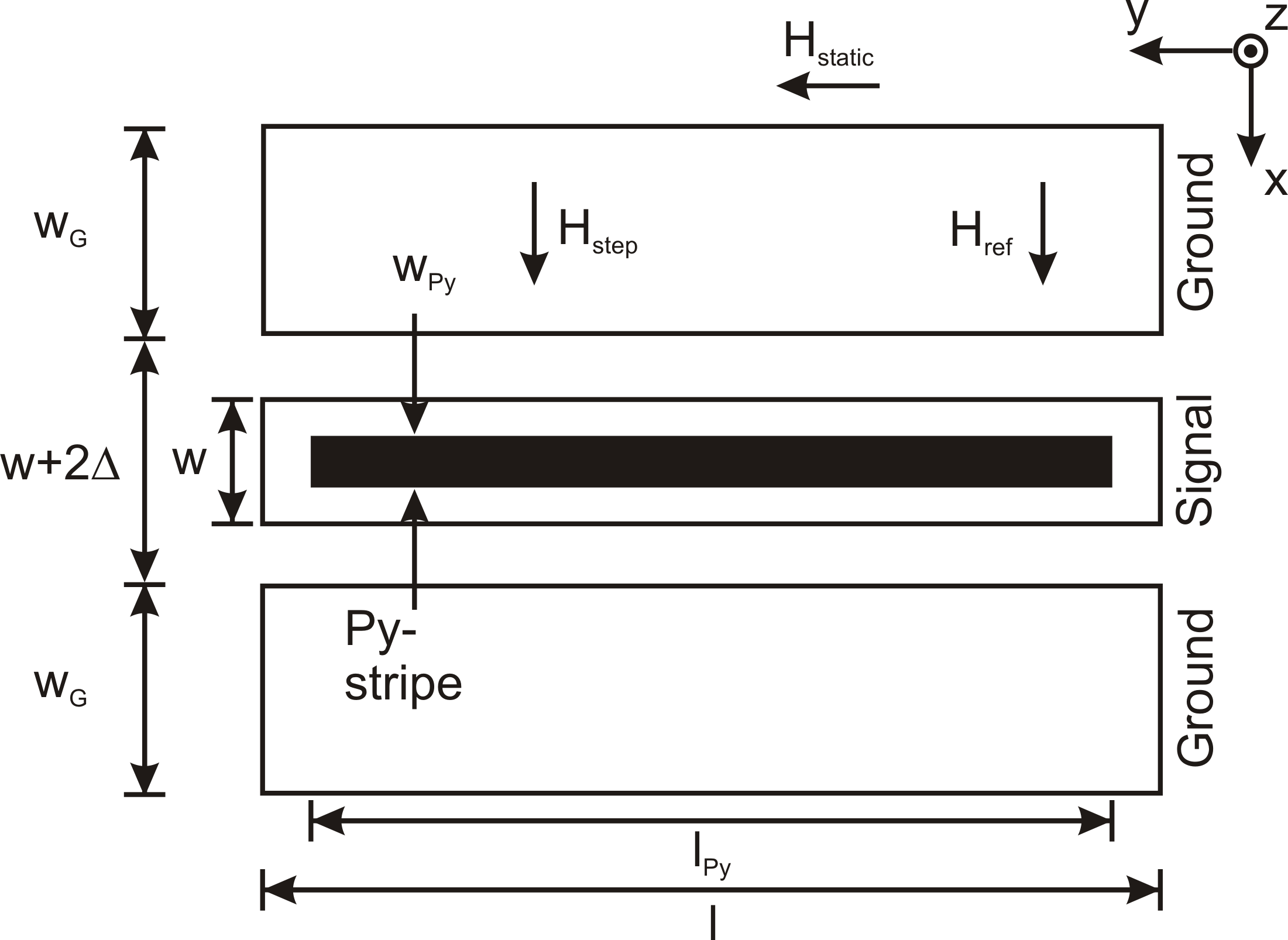}
\caption{Schematic view of the samples used in experiments (top view). The gap $\Delta$ between the $w_\mathrm{G}=300$\,$\mu$m wide ground contacts and the $w=64\,$$\mu$m wide center contact is $\Delta=13$\,$\mu$m. The length of the Py-stripe $l_\mathrm{Py}=0.75\cdot l$ where $l$ is the length of the CPW.}
\label{fig:samplegeometrie_hor}
\end{figure}

Two types of permalloy (Ni$_{81}$Fe$_{19}$) samples are studied, 20\,nm and 30\,nm thick respectively.
The sample design is shown in fig. \ref{fig:samplegeometrie_hor}.
After sputter deposition of the permalloy films, ion beam etching was used to pattern the rectangular permalloy in stripes. CPW consisting of 15\,nm titaniumoxid and 200\,nm gold, were subsequently lithographically patterned on top of each stripe such that they were aligned along the long dimension of the bars and centered on the stripes. The geometry of the coplanar waveguides was chosen for 50\,$\Omega$ impedance matching. The $w=64\,\mu$m wide center conductor is seperated by a gap of $\Delta=13\,\mu$m from the $w_\mathrm{G}=300\,\mu$m ground contacts. The length of the permalloy stipes is $l_\mathrm{Py}=0.75\cdot l$ where $l=0.75,1.0,1.5,2.0,3.0,4.5$mm is the length of the CPW for the 30\,nm thick thin films and $l=6\,$mm for the 20\,nm thick samples. The width of the 20\,nm thick samples was $w_\mathrm{Py}=10, 20, 30, 40, 50, 60, 64, 75$ and 85\,$\mu$m and for the 30\,nm samples $w_\mathrm{Py}=40, 50, 60, 70$ and 80\,$\mu$m.

Both time and frequency domain measurements were performed with an in-plane bias field applied along the long axis (y-axis) of the permalloy stripes. The amplitude of the bias field $\mu_0H_\mathrm{static}$ was varied in several steps from 0\,mT to 60\,mT. To derive the pure inductive signal resulting from magnetization precession an additional reference measurement was taken at a saturation field of $\mu_0H_\mathrm{ref}=60$\,mT applied along the x-direction. The precession signal was then derived by subtraction of the two traces. 

VNA-FMR transmission stray parameter $S_{21}$ measurements in the frequency domain were made over a frequency range from 40\,MHz to 30\,GHz at 0\,dBm power. An error correction due to a misalignment of the sample as described in reference \cite{Bilzer2007} was not necessary. Time-resolved PIMM data was acquired using a 20\,GHz sampling oscilloscope, applying a field pulse of 50\,ps rise time, 100\,ps duration and an amplitude of 1\,V \cite{Serrano-Guisan2008}. 

By fitting the time domain data to an exponential damped sinusoid and the frequency data to a symmetric and an asymmetric Lorentzian function we derived the effective damping parameter $\alpha$ and the resonance frequency $f$. Generally, the such derived $\alpha$ decreases with increasing applied field and saturates above a certain field threshold. At low fields $\alpha$ is extrinsically augmented by e.g. inhomogenieties. Therefore we use the asymptotic values of $\alpha$ at high fields as described in \cite{Counil2004} to derive the effective damping without contributions from inhomogeneous line broadening. Time and frequency domain measurements were cross validated by using two different measurement setups. Note, however that two magnon scattering processes inhibit the determination of the intrinsic Gilbert damping in in-plane magnetic field measurements and lead to an enhanced effective $\alpha$ as compared to measurements in out-of-plane field configurations \cite{Kalarickal2008,Landeros2008}.
\section{Results}
\begin{figure}[!t]
\centering
\includegraphics[width=1.0\linewidth]{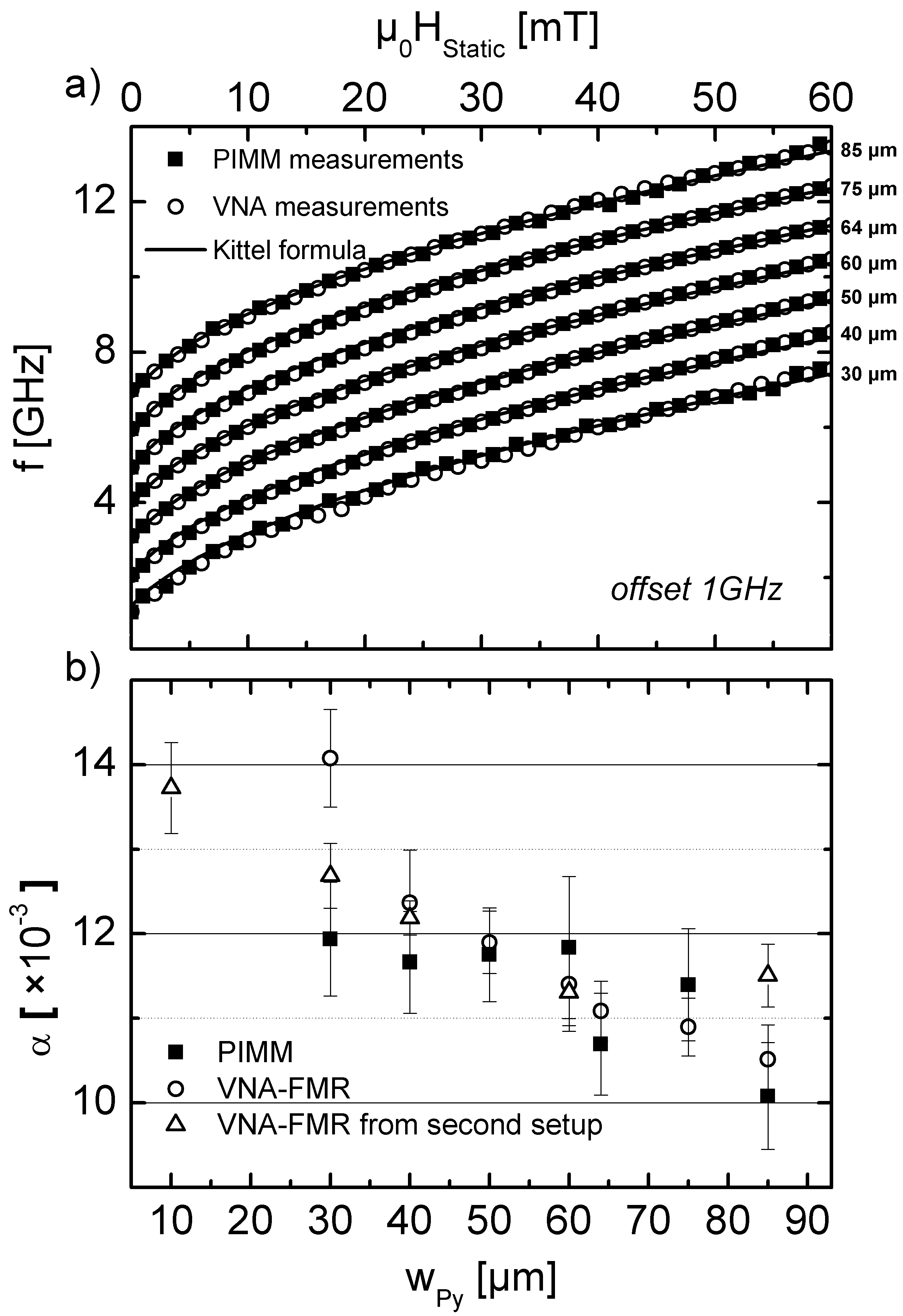}
\caption{The results of inductive measurements in time and frequency domain for the 20\,nm thick permalloy films.
a) The resonance frequency is plotted over the applied static field. The resonance frequencies are in good concordance with the Kittel formula (straight line) for both time (filled squares) and frequency domain (open circles) measurements. The data was shifted vertically for presentation. b) The width dependence of the effective damping for the 20\,nm thick samples. In both methods we observed a width dependence of the damping parameter. The damping decreases with increasing width of the elements. This result was cross validated by a second setup (open triangles).}
\label{fig:l6mm_frequ_damp_vs_width}
\end{figure}

Figure \ref{fig:l6mm_frequ_damp_vs_width}\,a) shows the precession frequency of the 20\,nm  thick films for different widths of the stripes derived from both time and frequency measurements.
They show a typical Kittel-FMR behaviour and can be superimposed regardless of sample width. The different geometries and the resulting different shape anisotropies do not significantly influence the precession frequencies. Also no shift in frequency was observed as reported in reference \cite{Counil2004,Guslienko2002, Schneider2004}. This is in good concordance with the similar demagnetizing factors for all elements calculated after \cite{Aharoni1998}. In contrast the measured effective damping $\alpha$ shows a decrease with increasing width of the permalloy stripes (fig. \ref{fig:l6mm_frequ_damp_vs_width}\,b). Here the damping obtained from different methods versus the width of the stripes is shown for the 20\,nm thick samples. A decrease of $\alpha$ with increasing width is observable. Note that this behaviour is consistently found for both time and frequency domain setups.

\begin{figure}[!t]
\centering
\includegraphics[width=1.0\linewidth]{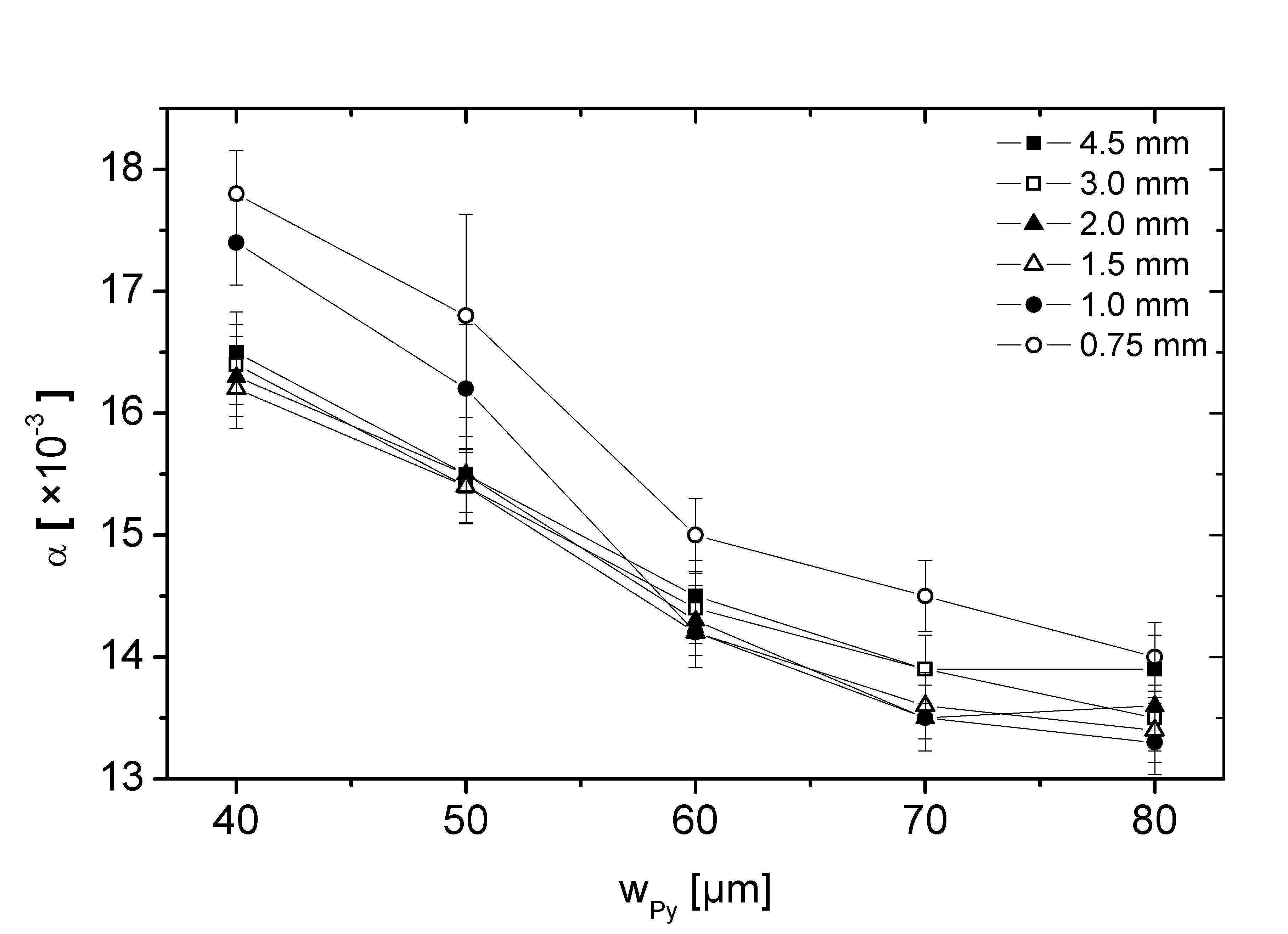}
\caption{The damping versus the width of the permalloy stripes for different length of the 30\,nm thick stripes. The decrease in damping  up to 20\,\% is clearly observable. For $w_\mathrm{Py}>$1\,mm a good agreement of the derived damping is found.}
\label{fig:30nm_frequ_damp_vs_width}
\end{figure}

As shown in figure \ref{fig:30nm_frequ_damp_vs_width} the same behaviour, as a function of sample width, is observed in the 30\,nm thin samples of the second type, patterned with different lengths and widths.
Again while the resonance frequencies can be well fitted by the Kittel formula (not shown) the damping becomes smaller with increasing width of the samples. We determined a similar change in damping of up to 20\,\% for all length and thicknesses. Also a small increase of $\alpha$ for the shortest sample is found. Note, however that for this sample the inductive signal is significantly smaller which could lead to an increased measurement uncertainty.
\section{Discussion}
The finite coplanar waveguide causes an in-plane inhomogeneity of the excitation field which excites spin waves with an efficiency P($\omega$) \cite{Kennewell2007}, where $\omega$ is the spin wave frequency. Counil et al. \cite{Counil2004} have shown that the excitation of spin waves due to the spatial inhomogeneity of the pulse field might modify the magnetic response and thus the derived value of $\alpha$. This result was validated by other groups \cite{Guslienko2002, Schneider2004}. Whereby in reference \cite{Schneider2004} the line broadening was studied as a function of sample thickness with respect to the CPW width and in reference \cite{Guslienko2002} as function of the wire aspect ratio $p$ ($p=$\,stripe thickness\,/\,stripe width). Following these arguments we would expect a more homogeneous pulse excitation and hence a \textit{lower} $\alpha$ for the narrower samples in contradiction to our experimental observations. Furthermore our measured data can not be consistently explained by standing spin wave modes \cite{Counil2004}. 

One possible origin of the decrease of $\alpha$ could be inhomogeneous precession near the edges of the Py stripes. This is in good concordance with the theoretical description of Guslienko et al. \cite{Guslienko2002} where they consider an effective pinning of magnetization at the lateral edges of the stripe. This pinning is related to the inhomogeneity of the internal dynamic field along the stripe width and is of dipolar nature. 

 With increasing width of the patterned stripes with respect to the center conductor width the edges of the stripes are moved into the gap of the CPW. Thus, their contribution to the measured inductive signal is drastically reduced. This is confirmed by measurements of a further set of test samples with the magnetic stripes situated within the gap. They show that the inductive signal from magnetic material situated in the gap can be practically neglected with respect to the magnetic material underneath the centre conductor.

These results have implications for the reliable determination of the intrinsic damping $\alpha$ based on patterned sample geometries. Magnetization dynamics in microstructured devices are modified with respect to extended unpatterned magnetic thin films. Note that in inductive magnetization dynamic measurements, as PIMM and VNA-FMR, the CPW behaves not only as emitter of the magnetic field pulse perturbation but also as the probe or antenna. It implies that, by these techniques magnetization dynamics are detected locally; i.e. only close to the CPW. Therefore, for microstripe devices with $w_\mathrm{Py}\le w$, one gets access to the magnetization dynamics of the whole structure including the edges. In this case the total effective damping parameter of the whole patterned structure will be measured which includes extrinsic contributions from the edges. However, for $w_\mathrm{Py}\ge w$, the contribution from the edges is not directly detected by the inductive experiment. Therefore, the influence of the edges on the measured magnetization dynamics is significantly reduced and the measured damping will approach the effective damping measured on extended thin films. 
Note again that the contribution of two-magnon scattering processes cannot be quantified from our experiments. The role of the sample geomentry on measurements of the intrinsic damping in out-of-plane fields \cite{Kalarickal2008,Landeros2008} will be the subject of future investigations.
 
\section{Conclusion}
We have shown that for inductive magnetization dynamics measurements on microstructured stripes with lateral dimensions comparable to the CPW dimension, the measured effective damping parameter $\alpha$ depends on the lateral size of the sample. A decrease of $\alpha$ up to 20\,\% is observed with increasing width. This effect is ascribed to an effective ``pinning'' of magnetization at the lateral edges of the stripe, inducing a dephasing of the magnetization precession and hence an enhanced $\alpha$. Therefore for accurate and reliable magnetization dynamic measurements by inductive techniques in microstructured ferromagnetic devices three important parameters must be taken into consideration: 1) CPW lateral dimension, 2) sample dimension, and 3) relative position and size of the ferromagnetic device with respect to the CPW. All these parameters can modify the resonance frequency and especially the derived effective damping parameter. 

\section*{Acknowledgment}
The research within this Euramet joint research project receives funding from the European Community's Seventh Framework Programme, ERA-NET Plus, under iMERA-Plus Project-Grant Agreement No. 217257.



%


\end{document}